\documentstyle[12pt,aasms4]{article}

\begin{document}

\title{HCG 16: A HIGH CONCENTRATION OF ACTIVE GALAXIES\\ IN THE NEARBY UNIVERSE}

\author{Andr\'e L. B. Ribeiro\altaffilmark{1}, Reinaldo R. de Carvalho\altaffilmark{2,4}}
\altaffiltext{1}{Divis\~ao de Astrof\'{\i}sica - INPE/MCT, C.P. 515 - 12201-970 - S. Jos\'e dos Campos, Brazil}

\author{Roger Coziol\altaffilmark{1}}
\altaffiltext{2}{Caltech - Astronomy Dept., Pasadena - CA 91125, USA}

\author{Hugo V.Capelato\altaffilmark{1}}

\and

\author{Stephen E. Zepf\altaffilmark{3,5}}
\altaffiltext{3}{Univ. of California - Astronomy Dept., Berkeley - CA 94720, USA}

\altaffiltext{4}{On Leave of absence from Observat\'orio Nacional - CNPq - DAF}

\altaffiltext{5}{Hubble Fellow}

\begin{abstract}

In the course of an extensive campaign to measure radial velocities
of galaxies in a selected sample of compact groups photometrically studied by
de Carvalho et al. (1994), we report the discovery of a system very rich 
in starburst galaxies and AGNs. This is the system HCG 16 of Hickson's 
(1982) catalog of CGs. The 7 brightest galaxies form a kinematical group
with bi-weighted estimate mean velocity of V$_{BI} = 3959 \pm 66$ km s$^{-1}$,
dispersion $\sigma_{BI} = 86 \pm 55$ km s$^{-1}$, a median radius 
$\langle {\rm R} \rangle  = 0.197$ Mpc, a mean density of $\langle {\rm D} 
\rangle = 217$ gal Mpc$^{-3}$ and a total absolute magnitude of 
M$_{B} = -22.1$. From their spectral characteristics, we have identified
one Seyfert 2 galaxy, two LINERs and three starburst galaxies.
Thus, HCG 16 appears to be a dense concentration of active galaxies.
In our sample of 17 Hickson groups, HCG 16 is 
unique in this regard, suggesting that it is an uncommon structure
in the nearby universe.

\end{abstract}

 
\section{Introduction}  

Compact groups (CGs) of galaxies may represent some of the densest
concentrations of galaxies known in the Universe and so may
provide ideal laboratories for studying the effects of strong
interactions on the morphology and stellar
content of galaxies. This concept has motivated several recent observational programs aimed at establishing
the dynamical reality of these structures
and signs of galaxy interactions. Whereas most of
these works (Rubin et al. 1991, Pildis et al. 1995) have mainly
considered objects in  which clear morphological signs of interactions
are evident,  studies made with larger samples have surprisingly shown
that the frequency of mergers in CGs is significantly less than that
predicted by simple dynamical arguments (Zepf  1993).
Confirming this result, in a search for tidal-tail induced dwarf
galaxies in 42 Hickson's (1982) CGs, Hunsberger et al. (1995) have
shown that only 7 of them exhibit clear signs of such objects.

Another question that may be addressed by studying CGs refers to the
environmental origin of the nuclear activity of galaxies (AGNs)
represented by the presence of nuclear emission lines that cannot be
explained in terms of normal stellar population. This longstanding question 
has been debated in the literature with no clear answer.  For instance,
whereas the studies  of Kennicut \& Keel (1984) and Keel et al.
(1985) have shown that the AGN phenomenom occurs more often in binary or
interacting systems, other
studies found no relevant correlations between nuclear activity and
interaction parameters (Dahari \& de Robertis 1988;
Laurikanen \& Salo 1995).  
However, it seems clear from studies of optically selected samples that
nearby AGNs avoid systems which are strongly interacting. 
Activity, if
present, seems rather due to intense starburst formation induced by the
interaction itself (Bushouse 1986).
This may not be true for the ultraluminous infrared galaxies 
which are mostly interacting galaxies and 
show an increasing probability of being Seyferts with increasing 
infrared luminosity (Veilleux et al., 1995). One may speculate that if
AGNs do really prefer interacting systems but avoid those that are strongly
interacting, then an ideal place to find them would be the CGs, or at
least a subclass of them which, although apparently very dense as
deduced from their radial velocities, show no signs of
violent interactions. 
 
In this {\sl Letter} we spectroscopically revisit one compact group which
has been previously noticed as presenting morphological signs of
interaction among its galaxies. This is the system HCG 16 of Hickson's
(1982) catalog of CGs. We have found this group
to be very rich in starburst galaxies and AGNs, being
possibly a very rare case of such a high concentration 
of active galaxies in such a dense environment.

\section {Data and Kinematic Properties}

HCG 16  and its neighbouring galaxies were spectroscopically observed in
the course of an extensive campaign aimed to measure radial velocities
of galaxies of a selected sample of CGs photometrically studied by
de Carvalho et al. (1994, hereafter dCAZ94).  
The spectra were taken at the 4m CTIO telescope, using the ARGUS
fiber feed spectrograph. The details of the instrumental setup and data
reduction are discussed  by de Carvalho et al. (1996).
Table 1
lists the galaxy number as given by dCAZ94 (column 1), 
positions R.A. and Dec. (columns
2 and 3), magnitudes in B (dCAZ94, column 3), and heliocentric velocities and
errors (columns 4 and 5). We list here only the seven galaxies defining
the group. A complete list of all the galaxies measured in the field is
presented by de Carvalho et al. (1996).

We have used the ROSTAT statistical package (Beers et al. 1990)
in order to analyse the velocity distribution of our
sample of galaxies around HCG16 (0.5$^{\circ}\times$0.5$^{\circ}$
around the center), which is complete down to B = 18.6$^{m}$. 
 The analysis revealed
the presence of a kinematical group consisting of the 7 brightest galaxies in
the field, with mean velocity (as given by the bi-weighted estimate) of
V$_{BI} = 3959\pm 66$ km s$^{-1}$ and dispersion $\sigma_{BI} = 86\pm
55$ km s$^{-1}$ (90\% confidence errors). A more detailed study is 
presented in Ribeiro et al.(1996a)

The compact group HCG 16 is  a
larger group than originally noted by Hickson, being composed of 7 galaxies.
In Figure 1, we present the distribution of galaxies in HCG 16. 
The original group
described by Hickson is composed
of 4 galaxies: 1, 2, 4 and  5. 
To these galaxies, we added 3 others, 3, 6 and 10. Galaxies
3 and 6 are quite luminous, which suggests that this new addition is of 
great importance for the dynamical structure of the group.     
The median of the projected separations of the galaxies in the group is
$\langle {\rm R}\rangle = 0.197$ Mpc (H$_{\circ} = 75~{\rm km~ s^{-1} Mpc}^{-1}$), and the mean 
density is $\langle{\rm D}\rangle = 217$ gal Mpc$^{-3}$. 
The density is therefore  smaller than the density $\langle {\rm D}\rangle = {10^4}$ 
gal Mpc$^{-3}$ originally determined by Hickson. While this density is 
lower than that in the central part of 
rich clusters of galaxies, HCG 16 has a density roughly 
30 times higher than those found for loose groups of galaxies, as determined 
by Maia et al. (1989). The total absolute magnitude of the system is 
M$_{B} = -22.1$ (as compared to M$_{B} = -21.5$ determined by Hickson).

\section {Spectral Properties}

In Figure 2, we present the spectra for the 6 emission--line galaxies
out of the 7 which constitute HCG 16. The seventh member of the group, 
galaxy 10, does not show any emission lines. 
Because the spectra are not flux calibrated, we
divided the number counts by their mean values, to compare the
relative intensity of the emission lines in the different galaxies. 
The galaxy 4 shows the most intense emission
lines. Except for the unusually high ratio of [NII]/H$\alpha$ and the
presence of a faint [OI]$\lambda$6300 line, this spectrum is very
similar to those of disk spiral HII regions. This galaxy is clearly
experiencing an intense starburst phase in or near its nucleus. 
Based on the similarity of the spectra, the galaxies 3, 5 and 6 are
also starburst galaxies, although at relatively lower intensities than
4. Galaxies 1 and 2 show a different type of spectrum, 
both with the high [NII]/H$\alpha$ ratio typical of AGNs.

The classification of the kind of activity encountered in the galaxies
of HCG 16 is based on the different line ratios shown by galaxies of
different activity classes. The criteria that we have used for our
spectral characterization are explained in detail in Ribeiro et al.
(1996b).  The presence or absence of a wide Balmer 
emission line component allows us to
distinguish 
between a Seyfert 1 and a Seyfert 2.  We distinguished between Seyfert
2 and LINERs based on the ratio [OIII]$\lambda$5007/H$\beta > 2.5$
(Coziol 1996). 
We adopted this definition  
because in many cases the [OII]$\lambda$3727 line, used by Heckman (1980)
to characterize the LINER type, was not available.  
For the starburst galaxies, we distinguished also
between HII galaxies and Starburst Nucleus Galaxies (SBNGs). This
distinction is based on a correlation between spectroscopic
characteristics and morphologies (Coziol et al. 1994).  In
general, the HII galaxies are high--excitation
([OIII]$\lambda$5007/H$\beta > 2.5$) small metal poor galaxies, while
the SBNGs are low--excitation massive and metal rich galaxies. Usually, 
the spectra of SBNGs indicate a mean excess of 0.2 dex in the
[NII]/H$\alpha$ ratio as compared to normal HII regions (Coziol et al. 1996).  

In Figure 3, we present the diagnostic diagram of
[OIII]$\lambda$5007/H$\beta$ vs. [NII]/H$\alpha$ for all the
emission--line galaxies in HCG 16. In this diagram, the dotted line
represents our criterion to distinguish between high and
low--excitation galaxies. The solid line is the empirical
separation established by Veilleux \& Osterbrock (1987) between
galaxies ionized by an AGN and galaxies ionized by stars. 
The uncertainties on the line ratios are determined based on
Poisson statistics. The high uncertainties in [OIII]$\lambda$5007/H$\beta$
for galaxies 1 and 2 reflect the weakness of the emission lines. In those 
galaxies for H$\beta$ the stellar absorption dominates over the nebular 
emission. Following our classification, HCG 16 contains 3 AGNs 
(2 LINERs and 1 Seyfert 2) and 3  starbursts. 
Based on the unusually high
intensity of the lines [SII]$\lambda\lambda$6716,6734 (see Fig. 2), galaxy 2 
looks more like a LINER than a Seyfert 2 (Rubin et al. 1991). The starburst
nature of galaxy 4 was already suggested by its high  emission in 
infrared (Sparks et al. 1986). The equivalent widths EW(H$\alpha$+[NII]) 
for the galaxies 3, 4, 5 and 6 are 44, 146, 50 and 10 \AA respectively. 
Except for galaxy 6, those values are comparable 
to those found in typical starburst galaxies (Kennicutt 1992).  

\section{X--Ray Characteristics}

X--ray emission from HCG 16 was first detected by Bahcall
et al. (1984). An analysis of ROSAT observations by Saracco \&  
Ciliegi (1995) suggests that in general the emission is point-like and
centered on the galaxies, although Ponman et al. (1996) have recently
presented evidence for diffuse intragroup X-ray emission from this group.
The two dominant galaxies, 1 and 2, appear to be in a common X-ray
envelope, for which Saracco \& Ciliegi give a total luminosity of
L$_x(0.5-2.3{\rm\ kev}) = 4.3\times10^{40}{\rm\ erg\ s^{-1}}$.
The fainter galaxies, 4 and 5
have resolved emission with luminosities $4.6\times10^{40}{\rm\ erg
s^{-1}}$ and $1.9\times10^{40}{\rm\ erg s^{-1}}$, respectively.
In the X-ray contour map presented by Saracco \& Ciliegi (1995) we found
a point-like emission at southeast which exactly coincides with galaxy
3. A crude estimate of the X-ray luminosity of this source gives $\sim
0.5\times10^{40}{\rm\ erg s^{-1}}$.

For comparison, mean values found by Green et al. (1992) give
L$_x(0.5-4.5 {\rm\ kev}) = 1.8\times10^{41}{\rm\ erg s^{-1}}$ for Seyfert 2,  
$2.2\times10^{40}{\rm\ erg s^{-1}}$ for LINERs and $7.0\times10^{39}{\rm\ erg s^{-1}}$
for starburst galaxies. In general therefore, the X--ray luminosities of the
galaxies in HCG 16 are comparable to those found in LINERs, and their X--ray emission intensities
decrease with the decreasing intensity of the starburst activity.

\section {Discussion}

HCG 16 is
a clear example of recent and multiple interacting
galaxies.   
As suggested by the spectroscopy, these interactions have
triggered a new phase of star formation in the galaxies 3, 4 and
5, while in the galaxies 1 and 2, it is likely that only the nuclei
were activated. The fact that 1 and 2 do not exhibit a starburst seems
to be in contradiction with the description of their morphologies by Rubin
et al. (1991) who reported signs of interactions in the forms of
weak antennas (in galaxy 1), tidail tails (galaxy 2), and a faint
bridge between 1 and 2. Indeed, following models of interacting galaxies, 
the development of such structures, especially the very long tails,
requires a few $10^8$ yrs (Barnes 1990), which is also comparable 
to the fading time scale of an induced starburst due to gas depletion 
(Mihos et al. 1993). 

Based on the absence of tidal features in the galaxies 3, 4
and 5 we could suppose that those galaxies are examples of a more recent
interaction (starbursts much younger than $10^8$ yrs) than the galaxies
1 and 2 which are at a more advanced stage of interaction.  
In favor of this scenario, we note that galaxies 1 and 2 may have
a common X-ray envelope.
The hot diffuse component could have been formed from gas that was 
stripped from the individual galaxies following their close encounter 
(Sulentic et al. 1995).
As judged from the remnant traces of interaction in the morphologies of 
these two AGNs, 
the galaxies 1 and 2 could have experienced a very strong starburst in their 
recent past (but surely older than few $10^8$ yrs). 
Since then, the burst has faded and only low activity in the
nuclei remains. More observations are needed to look for possible traces
of post--starburst activity near the nuclei of galaxies 1 and 2.
It is not clear however, if there
exists any direct relation between the starburst
event and the AGNs. In particular, galaxies 1 and 2 could
already have possessed a well--formed Black Hole in their nucleus which
was rejuvenated by a new infall of matter provided by the interaction.
	
A fundamental point we want to stress in this {\sl Letter} is
that our spectroscopic data support the identification of the
HCG 16 as a real compact group.
Both the kinematical parameters and the spectral
classification of the six brightest galaxies as active are
easily understood if HCG 16 is a bound, dense system.
In contrast, our observations are difficult to understand if the enhanced 
activity and density in HCGs is explained by projections along filaments 
incorporating a true pair (e.g. Hernquist, Katz, \& Weinberg 1995, Mamon 1995).
At the same time, HCG 16 is unique among our sample of 17 groups
as a clear case of compact structure, and groups like it may be
very rare, at least in the nearby universe (Ribeiro et al. 1996a).
The other structures have a wide variety of projections and dynamical 
configurations. This variety demonstrates how difficult it is to define 
small groups of galaxies, and how useful spectroscopy is 
for detecting groups both by picking out structures in velocity
space and by testing for nuclear activity.

\acknowledgments

We thank the anonymous referee for useful suggestions.
A.L.B. Ribeiro acknowledges the support
of the CNPq and 
R. Coziol acknowledges the financial
support of the Brazilian FAPESP, 
under contracts 94/3005--0.
S.E. Zepf acknowledges support from NASA through grant
number HF-1055.01-93A awarded by the Space Telescope Science Institute,
which is operated by the Association of Universities for Research in
Astronomy, Inc., for NASA under contract NAS5-26555.

\clearpage

\begin{deluxetable}{ccccrc}
\footnotesize
\tablecaption{Velocity distribution analysis of HCG 16 \label{tbl-1}}
\tablewidth{0pt}
\tablehead{
\colhead{Galaxy} & \colhead{R.A.} & \colhead{DEC} & \colhead{$B$} & 
\colhead{V} & \colhead{$\delta$V}\\ 
\colhead{} & \colhead{(1950)} & \colhead{(1950)} & \colhead{} & 
\colhead{(km s$^{-1}$)} & \colhead{(km s$^{-1}$)}
}
\startdata
  1 & 2 6 57.35 &   -10 22 20.1 &  12.88 &    \phn4073  & 10\nl 
  2 & 2 6 53.53 &   -10 22 10.1 &  13.35 &    \phn3864  & 10\nl 
  3 & 2 7 50.46 &   -10 33 24.8 &  13.35 &    \phn4001  & 13\nl 
  4 & 2 7 11.24 &   -10 22 58.2 &  13.61 &    \phn3859  & 13\nl 
  5 & 2 7 15.63 &   -10 25 11.6 &  13.66 &    \phn3934  & 13\nl
  6 & 2 6 38.78 &   -10 33 23.2 &  15.56 &    \phn3972  & 12\nl
 10 & 2 6 \phn9.30 &   -10 10 28.2 &  17.52 & \phn4000  & 31\nl
\enddata
\end{deluxetable}

\clearpage

\figcaption[Ribeiro.fig1.eps]{Position on the sky of the galaxies in HCG 16.
Each circle represents a galaxy, with the diameter being proportional to its magnitude (the greater
the brighter). The nearest 4 galaxies to the center form the original group, as defined by Hickson.
The 3 new galaxies are within a projected distance of 900 arcsec. Symbols are as follow: $\bullet$, galaxies
without measured radial velocity; $\otimes$, background galaxies;
$\odot$, galaxy at V = 3167 $\rm km s^{-1}$, which has been excluded
from the group  based on the consistency of the observed distribution
being drawn from a single gaussian parent population.}
\figcaption[Ribeiro.fig2.eps]{Optical spectra for the 6 emission--line galaxies
of HCG 16. For each spectrum, the counts were divided by the mean value
to facilitate the comparison between the galaxies. Each galaxy is identified by a running number
(see Table 1).
The letter in parenthesis refers to the original notation by Hickson (1982).} 
\figcaption[Ribeiro.fig3.eps]{Diagnostic diagram of emission line ratio and classification
of the different activity of the galaxies of HCG 16. 
The dotted line at log([OIII]/H$\beta$) = 0.4 establishes
a distinction between high--excitation and low--excitation galaxies (Coziol 1996). 
The solid line indicates 
the empirical separation between starburst galaxies
and AGNs as determined by Veilleux and Osterbrock (1987).}

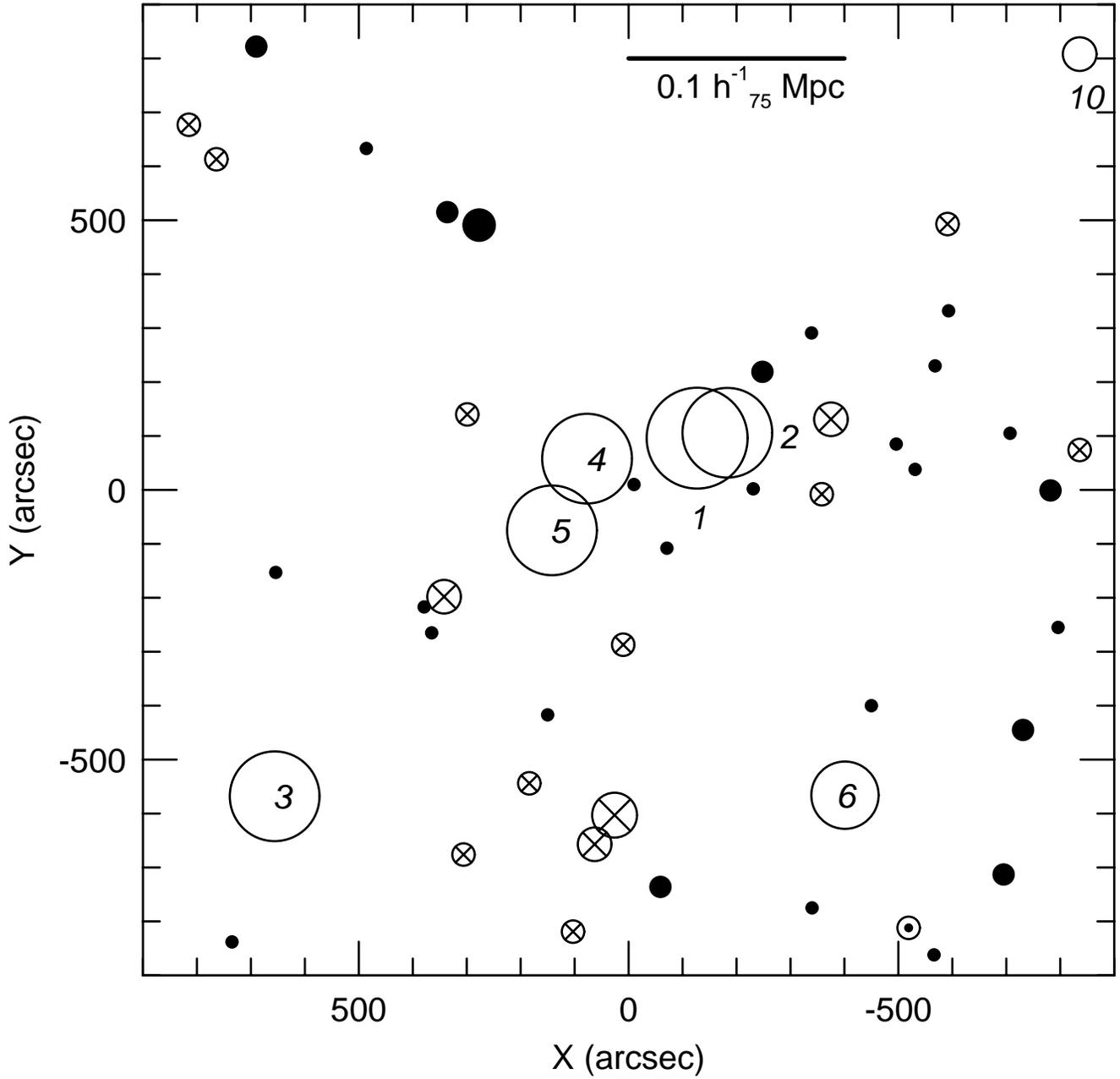
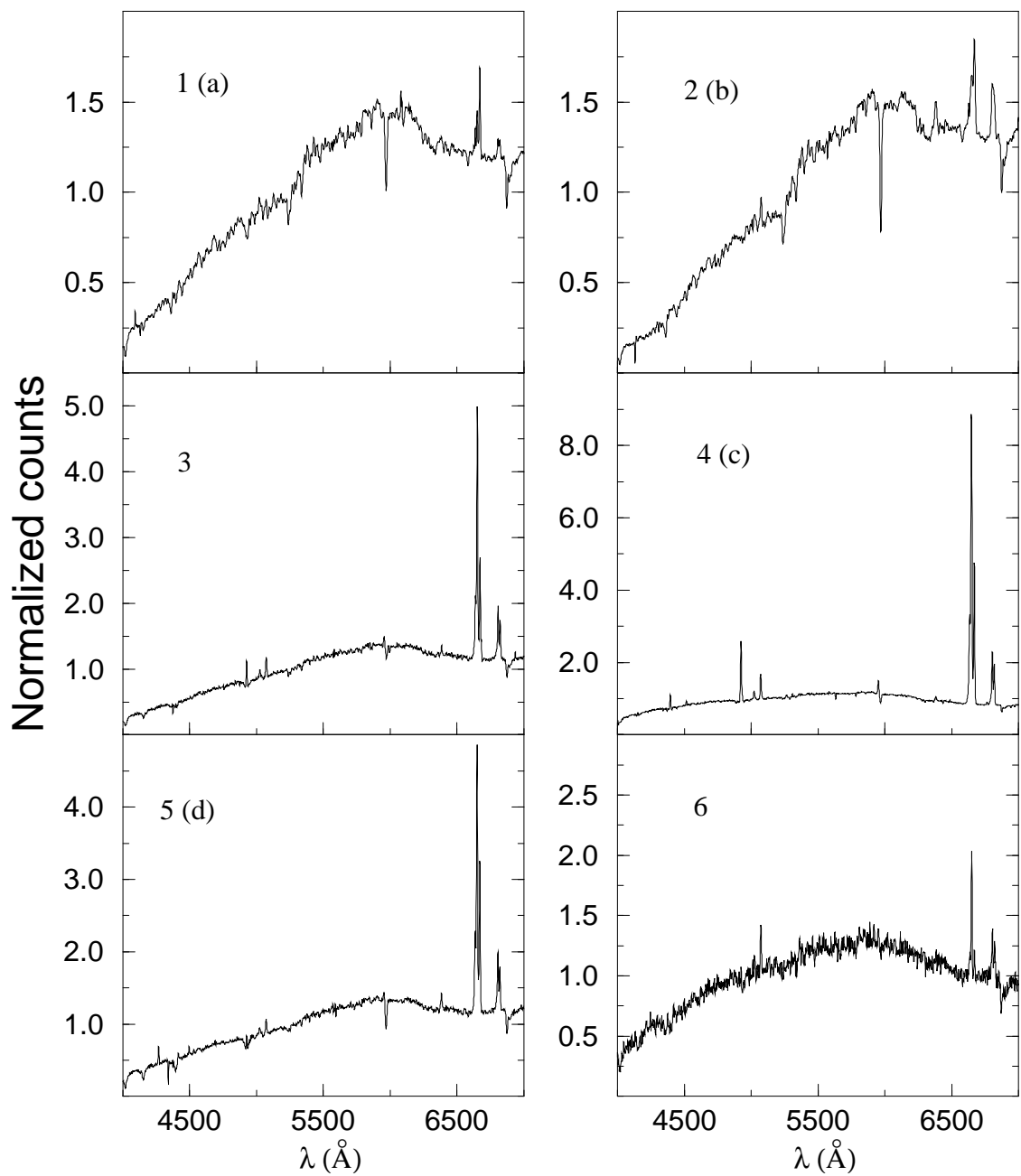
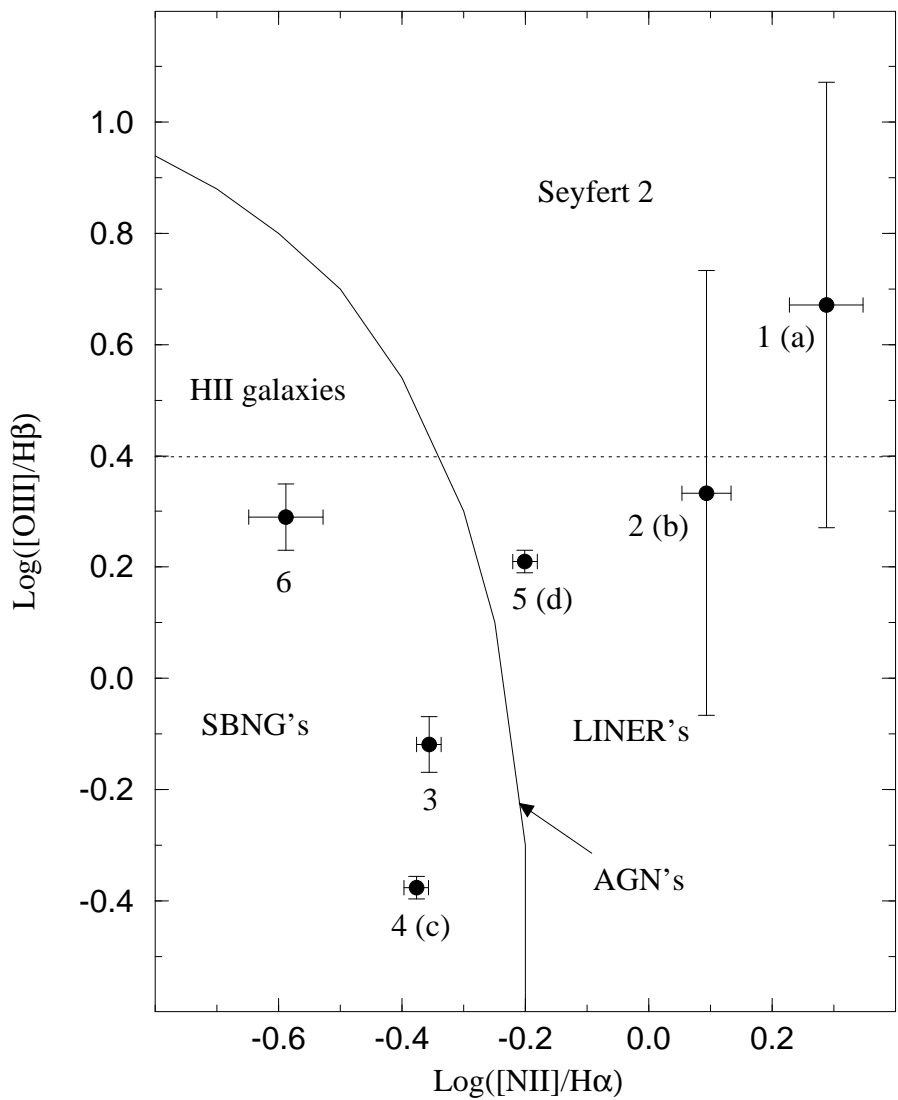
\end{document}